# Soft-Median Choice: An Automatic Feature Smoothing Method for Sound Event Detection

*Fengnian Zhao[1], Ruwei Li[1], Xin Liu[2], Liwen Xu[2]*

[1]Faculty of Information Technology, Beijing University of Technology, China
[2] Huawei Technologies Co., Ltd., China

zhaofengnian@emails.bjut.edu.cn, liruwei@bjut.edu.cn, liuxin159@huawei.com, xuliwen5@huawei.com

## Abstract

In Sound Event Detection (SED) systems, the lengths of median filters for post-processing have never been optimized during training due to several problems. No gradient is received by the lengths so they cannot be learned during back-propagation. The median-filtering inserted in the models also causes block in gradient flowing and the smoothing process misleads the model by ignoring errors. To resolve these problems, we provide different channels of features smoothed to different extents along with the original feature, so the model can optimize the weights while cognizing all the errors. We then use a linear layer to integrate the results and produce a linear combination. We further design the soft-median function to dredge the gradient flow. The proposed framework is called Soft-Median Choice (SMC). Experiments show that the SMC block not only automatically smooths the features based on the training set, but also forces the model to extract common features shared by all the frames of a sound event. The performance of the proposed method outperforms the baseline by over 10% of Event-Based F1 Score (EBFS) in both the validation and the evaluation set, and also slightly outperforms the single model of the state-of-the-art SED system.

## 1. Introduction

Sound event detection (SED) is a task to recognize the presence of sound events, and also detect their onsets and offsets. For the preparation of enough data for training, the intuitively perfect kind of data is the labelled data with time-stamps, the so-called *strongly labeled data*. However, labelling this kind of data by humans is costly and may result in human errors [1]. To reduce the dependence on strongly labeled data, recent models also use data with only the tag information, i.e., *weakly labeled data*, and also use *synthesized strongly labeled data* and *unlabeled data* [2]. With this heterogeneous dataset, many recent SED models use a backbone of CRNN to produce strong predictions of sound events[1][3][4]. To make use of weakly labeled data, the system also uses an aggregation method to converge the knowledge of frames and yield an integrated prediction of the whole audio clip. For the choice of aggregation methods, reference [5] compared the 5 existing aggregation functions with theoretic and practical evidence. Among them, linear softmax was proven to be the best. For the unlabeled data, learning this kind of data is conducted through semi-supervised learning[3][4][6][7]. With the use of the aforementioned kinds of data, the detection performance has been improved. However, as the size of model grows, the system is likely to suffer from over-fitting. With the help of data-augmentation, enough quantity of data is obtained for training[3][8][9]. This technique improves generalization performance.

The frame-level predictions may be non-consecutive, and may divide a single prediction of long period into multiple short predictions, causing a decline of the scores calculated based on correct predictions of onset and offset of a sound event. A common way for solving this problem is to apply a median-filtering for post-processing. The median filter slides a window through the signal and replaces the point in the middle with the median of the values within the window[10]. Due to the capability of smoothing the signal as well as preserving edges[11], median filtering is widely used in various fields to smooth the final result[12][13][14]. In the field of SED, median filtering is used to smooth the strong predictions of sound events and provides improvement.

The determination of the length of the window in median-filtering is conducted in two ways. One way is to determine empirically in advance, for example, a global length of a fixed period of time[2], or class-dependent lengths according to the different durations of sound events based on the statistics of the training set[15]. Nonetheless, the parameters chosen in advance are likely to be suboptimal, and the statistics in a particular set may be prejudiced and make the system difficult to adapt to real circumstances. The other way is to decide the lengths of median filters for different classes according to the result of the validation set[8]. This approach may gain huge improvement in the validation set. However, determining the hyperparameters based on a small dataset heighten the risk of overfitting.

For better performance of generalization, the lengths of the windows in median-filtering should be learned and optimized based on the training dataset. To reach this target, the median-filters should be utilized in the model for optimization during training. However, according to our knowledge, none of the pervious researchers in SED field has applied the median function as a part of the model[2][3][4][16][17]. Although the differentiable module of median blur with the backend of pytorch exists in kornia package[18], our study shows that a system with a differentiable median filter layer cannot converge well, resulting in poor performance.

The approach to solving this problem lies in a similar condition. The invention of softmax successfully resolved a similar problem. In the usages of multi-classification problems, the model must classify according to multiple categories. The function of one-hot argmax is used to get the final result where the category with the largest possibility gets the value 1 while the others get 0. Nevertheless, as a point with a value of zero does not contribute to the final result, according to the rule of back-propagation, the parameters that are related to this point cannot get any gradient, thus failing to optimize. To settle this issue, the softmax function, an approximation of argmax[19], is used after the classification layer. The softmax transforms all the elements of the vector into values between 0 and 1 with a sum of 1. The element previously with a value of absolute 0 now possesses a value for the convenience of back-propagation. This helps the parameters in the model to learn faster and converge well. Inspired by the invention of softmax, we design a function called soft-median, an approximation of arg-median. Replacing arg-median filters with soft-median filters enables all the parameters of the layers before the filter to get gradients, thus the model learns and converges faster.

In the best possible scenario, the size of soft-median filters should be a parameter that can be learned by back-propagation along with other parameters, whereas there is no such gradient that can be learned by the length of the window, not to mention that the lengths are integers; there is currently no way to learn the window length by back-propagation. Furthermore, the median-filtered results make the feature extractor to ignore the errors that are erased by the smoothing, which harms the performance. In order to allow the model to learn the parameters with convenience, while also preserving the unsmoothed result for learning from unignored errors, we use a series of soft-median filters to produce values smoothed to different extents, along with the unsmoothed result (or with a filter of length 1). We then acquire a linear combination of these results, allowing the model to learn the best weights for differently smoothed results. The whole structure is named as "Soft Median Choice" (SMC). Our proposed SED system based on the SMC achieves 48.6% and 0.694 of respectively event-based F1 Score (EBFS)[20] and Polyphonic Sound Detection Score (PSDS)[21] in the evaluation set.

The rest of the paper is structured as follows. Section 2 introduces the structure of the backbone of SED system based on CRNN, the SMC block and the design of Soft-median function as the main part of the SMC. Section 3 evaluates the performance of the proposed method with different configurations and discovers the reason for SMC's capability. The final conclusions are drawn in Section 4.

## 2. Proposed Method

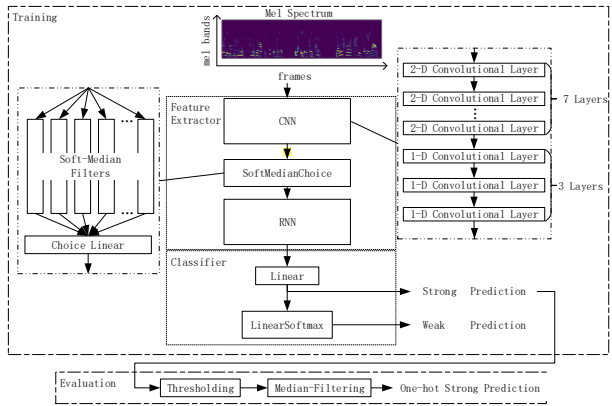

Figure 1: *An overview of the proposed method.*

Figure 1 is an overview of the proposed method. During the training process, the Mel Spectrum is first obtained after the input of an audio clip. A feature extractor based on CRNN and Soft-median Choice then extracts the features of every frame. Finally, the Classifier sorts the frames into different categories, and produces strong and weak predictions of sound events. During the evaluation process, the strong prediction goes through a post-processing of thresholding and median-filtering and results in the one-hot strong output.

### 2.1. SED Backbone Based on CRNN

The backbone of the SED system is based on a feature extractor of CRNN and a classifier including a linear layer and an aggregation function of *Linear Softmax*[5]. In the CRNN structure, the CNN is consisted of 7 layers of 2-D convolution and 3 layers of 1-D convolution conducted through the time axis. The following RNN block consists of 2 layers of Bidirectional Gated Recurrent Unit (BiGRU). The temporal structure of 1-D convolutions and RNN enables the model to grab more information through time, which is useful as sound signals are highly time-sequential. After latent presentations are obtained through the feature extractor, the classifier is equipped to produce final decisions. The linear layer with an activation of sigmoid classifies the sound events occurred in each time frame, and yields a strong prediction of strong labels with time-stamps. The following aggregation of Linear Softmax[5] produces a converged tag label of the whole audio clip.

### 2.2. Soft-Median Choice

Besides the backbone of the SED system, a novel structure named Soft-Median Choice is inserted between CNN and RNN. It is designed for the purpose of getting the knowledge of feature smoothed to different extents simultaneously. The diagram is shown in Figure 2.

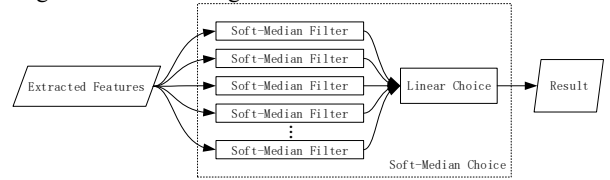

Figure 2: *The structure of Soft-Median Choice.*

The extracted feature is first processed by Soft-Median Filters of different window sizes, then a Linear Choice block integrates the results and yields a linear combination:

$$z = \sum_i w_i y_i + b \qquad (1)$$

where $y_i$ denotes the smoothed result of $i$-th Soft-Median Filter, $z$ is the output of the Linear Choice, $w_i$ is the weight of the $i$-th Soft-Median Filter and $b$ is the bias. The weights and bias are randomly initialized and learned by back-propagation. As no particular initialization method is designed for the linear layer in this structure[23][24], the bias is used for an additional supplementary learnable parameter to prevent the result yielded by the Linear Choice is too small at the beginning and influence the training process.

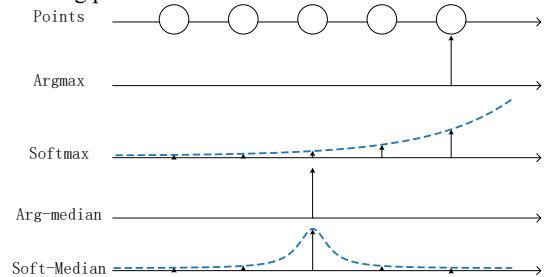

Figure 3: *An overview of different functions.*

*2.2.1. Soft-Median Function*

The Soft-Median Filter in Figure 2 is an approximation of median filters for the convenience of training the model. Soft-Median Function is used in the filters in the place of arg-median function. The convolutional arg-median function only picks the median point in a window, thus blocking the gradient flow. To solve this problem, we design a soft-median function and the reason and the approach of designing the soft-median is similar with that of Softmax. Figure 3 shows the distributions of 3 existing functions and the expected distribution of the Soft-Median.

As shown in Figure 3, the equation of Argmax can be denoted by:

$$W_k = \delta(x_k - x_{max}) \quad (2)$$

where $x$ is the symbol of data points, $W$ denotes weight, $k$ indicates the sequence number of the points within a window and $\delta$ denotes the unit impulse function.

Softmax utilizes the standard exponential function to each value and normalizes them by dividing the sum of the exponentials. The weights of Softmax are calculated by the following function:

$$W_k = \exp(x_k) / \sum_k \exp(x_k) \quad (3)$$

Similarly, the Arg-median function is denoted by:

$$W_k = \delta(x_k - \hat{x}) \quad (4)$$

where "$\hat{\ }$" represents the median of the elements within a window.

The Soft-Median function must satisfy three conditions: a) The weight must be the largest at the median point. b) The longer the distance between the point and the median is, the smaller the weight is. c) The weight must decrease sharply to zero so the function can be still approximated as a median function.

Due to the aforementioned reasons, the weight function of Soft-Median is chosen to be $x^{-2}$, but the weight of the median point itself cannot be 1, so we use a small $\varepsilon$ to divide some weights to other points, circumventing the median point to get all the weight. The equation is denoted as follows:

$$W_k = \left[(x_k - \hat{x})^2 + \varepsilon\right]^{-1} / \sum_k \left[(x_k - \hat{x})^2 + \varepsilon\right]^{-1} \quad (5)$$

It can be proved that when $\varepsilon \rightarrow 0$, (5) $\rightarrow$ (4). Note that each aggregated value is got by a weighted sum, shown in (6):

$$y_{ij} = \sum_k \left\{ \left[(x_k - \hat{x}_{ij})^2 + \varepsilon\right]^{-1} x_k / \sum_k \left[(x_k - \hat{x}_{ij})^2 + \varepsilon\right]^{-1} \right\} \quad (6)$$

where $j$ denotes the number of the points in the filtered result.

It should be mentioned that the denominator and numerator of the weights can be very large. For numerical stability, it is recommended that the weights are calculated first before multiplying $x_k$. The Soft-Median function successfully accelerates the training process and helps the model to converge to a better point, resulting in better performance.

## 3. Experiments and Results

### 3.1. Training Configurations

The datasets are from DCASE Task 4[2]. The number of samples in the training sets named "weak-real", "strong-synthetic" and "unlabeled-real" are respectively 1578, 2584 and 14412; the number of samples in validation and evaluation "strong-real" datasets are 1168 and 692. Each audio sample is of a length of 10 seconds. The models all use a mean-teacher method[22] for semi-supervised training. The data augmentation of time-shifting and mix-up is applied during training[3]. The performance indicators are EBFS[20] and PSDS[21]. They suggest better performance when they are larger. As EBFS is for ranking in DCASE challenges, the evaluation of performance is mainly based on EBFS with PSDS as a supplementary indicator.

### 3.2. Evaluations of the Proposed Method

*3.2.1. Effects of Model Architecture*

In this section, several designs of model architectures are discussed. In order to get rid of the influence of other settings, all of the experiments are done in the validation set without the median-filter for postprocessing, with the initialization of He with a gain of $\sqrt{5}$.

Firstly, the object for the SMC block is determined. The SMC block can be utilized to various objects. The first smoothing approach is applying the block to the predicted probabilities and assign each class an SMC block, in the hope that each different class can choose a different length of median-filter. The second smoothing approach is applying one block to the predicted probabilities, so that the model can optimize the length globally. The third approach is applying the SMC block to framewise latent presentations produced by the CNN. The last approach is applying the SMC block to the latent presentations produced by the RNN.

According to Table 1, using the SMC block with predicted probabilities is not a good idea. The proposed block is not a simple post-processing method, but a structure inserted to the model itself and greatly affects the training process. Although directly applying this part to the probabilities will make the model get better results, the difference between the results and ground truth is ignored, thus the feature extractor cannot learn enough information to extract better latent presentations. This is possibly the real reason why the median filter post-processing is not used during training. From the evaluation we can infer that the SMC applied to latent presentations of CNN not only automatically smooths the features, but also forces the CNN to extract robust features against the smoothing, which is exactly the crucial features that are commonly processed over the frames of a sound event.

Secondly, the lengths of the Soft-median filters of the SMC are chosen. Two schemes are evaluated. For the lengths, $1 \leq length < 60$. In the first scheme, the gap is 2, while in the second scheme, the gap is 4. Therefore, the number of filters are 30 and 15 respectively. An additional experiment of only using one filter of 7 points is also conducted. As shown in Table 2, the scores of 15 filters are the highest. That is because the difference between the smoothed features of lengths with a gap of 2 is little, but the whole model becomes more complicated, which affects the performance. The model with only one filter fails to know the errors erased by median filters, also resulting in poor performance.

Finally, Table 3 shows that the Choice Linear is better with bias versus without bias. As the Choice Linear integrates large quantities of data into small quantities of data, during the beginning of training, the CNN is short of enough gradient to learn from. The initial bias faster the training process and improves the performance.

Table 1: *Effects of the object of using Soft-Median Choice.*

| Object | EBFS [%] | PSDS |
|---|---|---|
| Predicted Probabilities (Class-dependent) | 30.0 | 0.548 |
| Predicted Probabilities (Global) | 30.3 | 0.599 |
| Latent Presentations (CNN) | **40.6** | **0.624** |
| Latent Presentations (RNN) | 30.4 | 0.591 |

Table 2: *Effects of the filter numbers of using Soft-Median Choice.*

| Number of Soft-Median Filters | EBFS [%] | PSDS |
|---|---|---|
| 30 (from 1 to 59) | 40.3 | 0.622 |
| 15 (from 1 to 57) | **40.6** | **0.624** |
| 1 (only 7) | 33.9 | 0.593 |

Table 3: *Effects of Choice Linear with/without bias.*

| Bias | EBFS [%] | PSDS |
|---|---|---|
| Without Bias | 36.7 | 0.609 |
| With Bias | **40.6** | **0.624** |

### 3.2.2. Effects of Initialization

Xavier and He have already discussed the way of initialization after most symmetric activations[23] and asymmetric activations[24]. However, the initialization after the Soft-Median filters is not previously discussed and designed as yet, thus allowing us to attempt several ways of initialization to the weights. The approaches are Xavier and He uniform initializations with the gains shown in Table 4. The bias is initialized uniformly according to the default of the linear layer in pytorch[25]. According to the results, the He uniform initialization with a gain of $\sqrt{5}$ (the default initialization of the linear layer in pytorch) is the best.

Table 4: *Effects of the Initialization of Choice Linear.*

| Initialization | Gain | EBFS [%] | PSDS |
|---|---|---|---|
| Xavier | $\sqrt{2}$ | 35.6 | 0.593 |
| | $\sqrt{3}$ | 35.9 | 0.607 |
| He | $\sqrt{5}$ | **40.6** | **0.624** |
| | $\sqrt{6}$ | 37.5 | 0.618 |

### 3.2.3. Effects of Post-processing

The post-processing of thresholding and median-filtering shown in Figure 1 is for yielding one-hot and smoother framewise predictions. Two kinds of post-processing are evaluated based on the CRNN Backbone (CB) introduced in Sec 2.1. The Global Post-Processing (GPP) uses a global threshold of 0.5 and a global size of median-filter of 0.45s /7 points. The Class-Dependent Post-Processing (CDPP) searches for class-dependent parameters of the optimal thresholds based on the validation set from 0.1 to 0.9 with a gap of 0.1, and the optimal sizes of the median-filter from 1 to 61 with a gap of 2. If none of them are present, only a thresholding of 0.5 is used.

Table 5 shows the effects of different kinds of post-processing. On the one hand, the GPP brings an improvement between 1% and 2% of EBFS in the validation set, with or without the SMC. Its failure of improvement in the evaluation set, however, shows that the empirical parameters may do not fit the distributions of the evaluation set. On the other hand, the CDPP brings improvement both in the validation set and the evaluation set. For the model architecture, the model with SMC outperforms the model without SMC under identical circumstances. The method of CB+SMC+CDPP is the best method and is thus the ultimate proposed method of this paper.

Table 5: *Effects of post-processing.*

| Method | Validation | | Public Evaluation | |
|---|---|---|---|---|
| | EBFS [%] | PSDS | EBFS [%] | PSDS |
| CB | 38.5 | 0.623 | 43.2 | 0.651 |
| CB+GPP | 39.8 | 0.648 | 43.1 | 0.667 |
| CB+CDPP | 44.2 | 0.669 | 44.6 | 0.680 |
| CB+SMC | 40.6 | 0.624 | 47.6 | 0.669 |
| CB+SMC+GPP | 42.4 | 0.648 | 47.3 | 0.684 |
| CB+SMC+CDPP | **46.2** | **0.664** | **48.6** | **0.694** |

### 3.2.4. Comparison with Other Systems

Table 6 shows a comparison of the proposed method with the baseline and the single model of the winning system of [26] in Task 4 of DCASE 2020. The performance of our model outperforms the baseline by over 10% of the EBFS in both datasets and slightly outperforms the winning system. Compared with the winning model that utilizes novel models borrowed from other fields and gains improvement in the SED system, our work applies a tiny framework to the conventional CRNN network and shows that there still is room to ameliorate based on the simple kind of CRNN network.

Table 6: *Comparison with representative models in TASK 4 of DCASE 2020*

| Method | Validation | | Public Evaluation | |
|---|---|---|---|---|
| | EBFS [%] | PSDS | EBFS [%] | PSDS |
| Baseline (2020)[2] | 34.8 | 0.610 | 38.1 | 0.552 |
| Miyazaki[26] | 46.0 | 0.643 | - | - |
| CB+SMC+CDPP | **46.2** | **0.664** | **48.6** | **0.694** |

## 4. Conclusions

In this paper, an SED system based on SMC is proposed. The latent presentations extracted by CNN is smoothed by the Soft-Median filters with variable lengths and a linear combination of them is automatically learned for RNN and classification. The performance of our system outperforms the baseline with 11.39% and 10.5% of EBFS respectively in the validation set and the evaluation set. It also outperforms the state-of-the-art system and with low complexity. Our work inspires more study to be done based on the Soft-median function or the SMC block.

Beside the role of automatic smoothing similar to post-processing, the SMC influences the process of training, which is presumably the reason why it brings improvement. In our opinion, the proposed SMC block not only smooths the features automatically, but also gives the model a glimpse the truly meaning of sound events rather than only knowing separate frames. Our work may draw deep thoughts of how to allow the model to know more about the conception of sound events.

## 5. Acknowledgments

This work was supported by the National Natural Science Foundation of China (No. 61971016) and Beijing Municipal Education Commission Cooperation Beijing Natural Science Foundation (No. KZ201910005007).